\documentclass[preprint]{aastex}

\slugcomment{Submitted to Meteorit. Planet. Sci., October 2001}

\shorttitle{Injection of Radioactivities}
\shortauthors{Vanhala & Boss}

\begin{document}

\title{Injection of Radioactivities into the Forming Solar System}

\author{Harri A. T. Vanhala}
\affil{Dept. of Physics and Astronomy, Arizona State University, \\
       PO Box 871504, Tempe AZ 85287-1504}
\email{harri.vanhala@asu.edu}

\and

\author{Alan P. Boss}
\affil{Department of Terrestrial Magnetism, Carnegie Institution of
       Washington, \\
       5241 Broad Branch Road NW, Washington, DC 20015-1305}
\email{boss@dtm.ciw.edu}

\begin{abstract}
Meteorite studies have revealed the presence
of short-lived radioactivities in the early solar system.
The current data suggests that
the origin of at least some of the radioactivities requires
contribution from recent nucleosynthesis at a stellar site.
This sets a strict time limit on the time available for
the formation of the solar system
and argues for the theory of the triggered origin of the solar system.
According to this scenario,
the formation of our planetary system was initiated by the impact of
an interstellar shock wave on a molecular cloud core.
The shock wave originated from a nearby explosive stellar event
and carried with it radioactivities produced in the stellar source.
In addition to triggering the collapse of the molecular cloud core,
the shock wave also deposited some of the freshly synthesized
radioactivities into the collapsing system.
The radioactivities were then incorporated into the first solar
system solids, in this manner leaving a record of the event in
the meteoritic material.
The viability of the scenario can be investigated
through numerical simulations
studying the processes involved in mixing shock wave material
into the collapsing system.
The high-resolution calculations presented here
show that injection occurs through
Rayleigh-Taylor instabilities, the injection efficiency is approximately 10\%,
and temporal and spatial heterogeneities in the abundances of the
radioactivities existed at the time of their arrival in
the forming solar system.
\end{abstract}


\section{Introduction}

Studies of meteoritic material have revealed the presence of short-lived
radioactivities in the early solar system
\citep{was85,cam93,pod97,gos00,mck00}.
The origin of these dozen or so confirmed
($^{41}$Ca, $^{26}$Al, $^{60}$Fe, $^{10}$Be, $^{53}$Mn,
$^{107}$Pd, $^{182}$Hf, $^{129}$I, and $^{244}$Pu)
or suspected ($^{99}$Tc, $^{36}$Cl, $^{205}$Pb, and $^{92}$Nb) short-lived
radionuclides has been a subject of intense investigation
over the last few years.
There are two basic ways to explain the production of the radioactivities:
via stellar nucleosynthesis or through local production in
the early solar system.

A stellar nucleosynthetic source (a supernova or an AGB star) has been
the leading explanation for most of the nuclei
\citep{cam93,cam01a,was94,was95,was98}.
This source works well for the longer-lived radioactivities, because
there is plenty of time to have them produced in stellar interiors and
then mixed into the interstellar medium, including the molecular
cloud core from which the solar system was formed.
However, the case is more complicated for the shortest-lived
radionuclides - mean life less than a few million years - since
their presence sets a time limit of one million years or less
for the time available between their production
and their incorporation in the solar system material.
This has led to the formulation of the idea of the triggered origin of
the solar system \citep{cam77,bos95,cam97,vab98,bos00},
which suggests that the formation of the solar system was initiated
when an interstellar shock wave propagating from a nearby explosive
stellar event impacted on a molecular cloud core.
In addition to triggering the collapse of the core earlier than it
would have occurred otherwise, the shock wave
also enriched it with freshly synthesized radioactivities
produced at the stellar site and carried by the shock wave.

Another possibility for the origin of the radionuclides
is to have them produced locally in the solar nebula.
According to this scenario, the radionuclides
were produced during spallation reactions involving energetic particles
emanating from protosolar flares \citep{shu97}.
Since the radioactivities are produced locally,
there are no time constraints for the formation of the solar system,
and therefore no need to connect the origin of the solar system
with a nearby explosive stellar event, as long as all the detected
shortest-lived radioactivities can be produced through this mechanism.

The current meteorite data suggests that the best explanation
for the origin of the radioactivities requires both scenarios.
Local irradiation models appear to have difficulty in matching the observed
abundance ratios of the nuclides, especially that of $^{26}$Al to $^{41}$Ca
\citep{sri96,sah98,lee98}.  Even if it may be possible to overcome this 
problem by assuming shielding of the refractory inclusions by a less refractory
mantle \citep{gou01},  local production scenarios cannot account for $^{60}$Fe,
and it therefore requires a stellar nucleosynthetic source \citep{gos01}.
The detection of the short-lived isotope $^{10}$Be in an Allende inclusion
\citep{mck00}, has stirred further debate, since the nuclide is thought
to be produced only by nuclear spallation reactions.  Therefore,
its existence has been used to argue in favor of local irradiation scenarios
\citep{mck00,gou01}, but it is unclear
whether the radioactivity was produced in the solar nebula,
or in an earlier phase of evolution, such as
in expanding supernova envelopes \citep{cam01b} or
in the winds ejected from H-depleted Wolf-Rayet (WR) stars \citep{arn00}.
These considerations lead to the conclusion that
a combination of stellar nucleosynthesis coupled with stellar and/or
local irradiation appears to be the best explanation for
the known isotopic anomalies.
Indeed, there is accumulating evidence
that the presence of the short-lived radioactivities
in the early solar system may require a fairly involved history
for their complete explanation \citep{mey00}.

The theory of the triggered origin of the solar system therefore
remains an attractive scenario to explain the presence of at least
some of the short-lived radioactivities in the early solar system.
In the last few years, the viability of the proposal has been
investigated through numerical simulations using
several different simulation methods
\citep{bos95,fos96,fos97,bos97,bos98,cam97,van98,van00};
for reviews, see \citet{vab98} and \citet{bos00}.
The simulations suggest that molecular cloud cores can be triggered into
collapse by moderately slow ($\sim$10-45 km s$^{-1}$) shock waves
\citep{bos95,fos96,cam97,van98},
and the time scale of the process, $\sim$10$^5$ yr
\citep{fos96,van98}, is sufficiently
short for the radioactivities to have survived in the measured amounts.
Calculations studying the mixing of radioactivities
into the forming solar system have shown that shock wave material can be
injected into the collapsing system when the postshock gas cools rapidly,
resulting in (nearly) isothermal shocks \citep{fos97,van98,van00}.
In this case, shock wave material
is injected into the collapsing molecular cloud
core through Rayleigh-Taylor (RT) fingers, with an efficiency of 10-20\%
and over a time period of 700,000 years \citep{fos97,van00}.
The radioactivities can be injected into the collapsing
system even if they are far behind the leading edge of the shock wave
\citep{bos98}.
It is also possible that shock wave material can be injected into
the system in non-isothermal shock waves where
the postshock gas remains hot behind the shock front, but
these aspects of the problem have not been investigated further
because they appear to be beyond current computational power \citep{van98}.
Consequently, the discussion of the injection process
has concentrated on the isothermal case.

Previous calculations have suggested
that the abundances of the injected material may have experienced
spatial and temporal variation in the early solar system.
This is based on two discoveries of the interaction between the shock wave
and the molecular cloud core.
First, there appears be a lag of a few $\times$ 10,000 yr
between the time the center of the compressed molecular
cloud core is pushed into collapse and the time the radioactivities carried
by the shock wave arrive deep in the system \citep{van00}.
This is due to the fact that while the core can be pushed into collapse
by the compressional wave transmitted through the core as the shock wave
is decelerated by the outer layers of the core,
injection becomes efficient only after
the RT-fingers have developed fully at the surface of the core.
This makes it possible for the amount of radioactivities to have varied
at different times within the forming solar system.
Second, the calculations suggest
that the RT-fingers remain well defined down to the level of a few
tens of AU, making it possible for spatial heterogeneities to exist
in the matter falling into different parts of the protostellar disk
\citep{van00}.
However, the resolution of the previous calculations has been
insufficient to make these suggestions conclusive.

The possibility of heterogeneities in the early solar system
has received support from meteorite studies.
For example, the absence of evidence of live $^{26}$Al in some
refractory inclusions might imply their
formation in the nebula
just prior to the arrival of the freshly-synthesized $^{26}$Al deposited
by the shock wave \citep{sah98}.
It is also possible that these inclusions,
as well as chondrules, which in contrast to typically
$^{26}$Al-enriched Ca,Al-rich inclusions (CAIs), show
lower values for $^{26}$Al enrichment or none at all
\citep{hut94,hut95,rus96,kit00,hus01},
formed in nebular regions largely devoid of $^{26}$Al, whereas
the CAIs were formed out of $^{26}$Al-rich material
\citep{mac95}.
Spatial heterogeneity is also implied
by the strong evidence for a radial gradient in the distribution
of $^{53}$Mn in the solar nebula \citep{lug98}.
There are some arguments
(\citet{mac95,nic99}; see the discussion by \citet{pod94})
that suggest that the heterogeneity may have been temporal rather than spatial,
but this is currently uncertain, and more data is required.

In order to address the details of the injection process, such as
the issue of the origin and longevity of heterogeneities,
we have initiated a study of the interaction between the molecular cloud core
and the shock wave at high spatial resolution \citep{van00}.
The work described in this paper is the latest in a series of calculations
studying the injection process under isothermal conditions.
The resolution of the current study is sufficiently high
to describe the behavior of the injected material
at the time it arrives to the forming solar system and determine whether
the heterogeneities discovered by the earlier calculations persist
at this stage.
The system studied in the current calculations is described in {\S} 2,
and the results of our study are given in {\S} 3.
In {\S} 4 we summarize the results and briefly discuss
their implications.

\section{Simulation method and the initial system}

The study presented in this paper
is a continuation of the calculations of \citet{van00}.
The two-dimensional VH-1 hydrodynamics code,
which is based on the piecewise-parabolic method (PPM)
and includes the effect of the self-gravity of the gas,
is described in greater detail by \citet{fos96}.
Two complementary methods - a color field and tracer particles -
are used to follow the behavior of the shock flow material
as it impacts the molecular cloud core \citep{fos97}.

The initial conditions of our calculations are
the same as the standard cases of \citet{fos97} and \citet{van00}:
a marginally stable Bonnor-Ebert sphere joining smoothly
to the surrounding medium.
The cloud initially has a radius of 0.058 pc, temperature of 10 K,
central density of $\rho_{\rm c}$ = 6.2 $\times$ 10$^{-19}$ g cm$^{-3}$,
and contains one solar mass of material.
The intercloud medium has
$T_{\rm icm}$ = 10 K and $\rho_{\rm icm}$ = 3.6 $\times$ 10$^{-22}$ g cm$^{-3}$.
The shock wave is represented by a top-hat model,
in which the edge of the wave (thickness
0.003 pc) is given the velocity $v_{\rm edge}$ = 20 km s$^{-1}$, density
$\rho_{\rm edge}$ = 3.6 $\times$ 10$^{-20}$ g cm$^{-3}$
and temperature $T_{\rm edge}$ = 10 K,
while the wind behind the leading edge has
$\rho_{\rm wind}$ = 3.6 $\times$ 10$^{-22}$ g cm$^{-3}$, $T_{\rm wind}$ = 10 K
and $v_{\rm wind}$ = 0 km s$^{-1}$.
The cylindrical coordinate ($r$, $z$) grid is taken to be axisymmetric
around the $z$ axis and extends
to $r$ = 0.088 pc and -0.176 pc $< z <$ 0.088 pc,
with the shock wave approaching from the $+z$ direction.

In accordance with the earlier studies, the wind is initially at rest.
In general, the wind would be expected to have a non-zero velocity.
However, we chose to keep the initial conditions the same as in
previous calculations to make the calculations comparable.
Also, previous calculations have indicated the
the injection results do not change appreciably from the standard case
with different wind velocities \citep{fos97,bos98}.

To further keep the current calculations comparable with previous studies,
we use an isothermal equation of state, with
the adiabatic index $\gamma$ = 1.00001 (see discussion by \citet{fos97}).
Since we are interested in investigating the injection process under
optimal conditions, isothermal conditions are an appropriate choice
\citep{fos97,van98,van00}.

In the current calculations, the two-dimensional uniform grid has
a resolution of 960 x 2880, twice the number of zones in each dimension
as in the highest-resolution calculation of \citet{van00}, and
16 times the number of zones in each dimension of the original case
of \citet{fos97}.
The size of one zone in these calculations is
19 AU (9.2$\times$10$^{-5}$ pc),
not enough to discuss the distribution of injected material at the scale of
the present-day solar system, but sufficient to describe the
injection process at the time of the formation of the solar system.

\section{Results}

The results of our simulations are shown in {\uppercase {Figs.}} 1-3.
The basic results of our calculations follow well the pattern
described by \citet{fos96,fos97} and \citet{van00}.
When the shock wave impacts the molecular cloud core,
it is stalled at the facing side of the core, while material
at the sides can sweep past the core.  A compressional wave is transmitted
through the core as the shock wave is decelerated by the outer layers of
the cloud, and eventually the central parts of the system are pushed
into collapse.  At this point the Courant condition is violated
due to the gravity becoming very strong and the calculation stops.
The evolution of the compressed core is characterized by
an initial growth spurt - about 0.5 solar masses over $\sim$100,000 years -
after which the growth of the protostellar core slows down until
the final protostellar mass is reached.


While the core is being compressed by the shock wave, Rayleigh-Taylor
instabilities develop at the surface of the core ({\uppercase {Fig.}} 1).
Shock wave material collects in clumps at the surface of
the compressed core, and RT-instabilities developing at these locations
inject shocked material into the collapsing system.
The complex structure of the RT-fingers is evident in {\uppercase {Fig.}} 2,
which shows a closeup of the last four panels of {\uppercase {Fig.}} 1.
The RT-fingers develop at different times at different parts of the core,
according to the time at which they first come into contact with
the shock wave: the facing side of the core has the instabilities
develop first, while the fingers at the sides of the core are just
beginning to reach deep into the system at the time the calculation stops.
The injection efficiency - the amount of shock wave material captured
by the collapsing system with respect to that originally incident on
the molecular cloud core - is approximately 10\%, the same as in
previous calculations.


Our calculations show that the injection becomes efficient only after
the RT-fingers have developed fully at the surface of the compressed core.
Consequently, there is a lag between the time
when the central regions of the core are significantly compressed
and when the shock wave material arrives in the inner regions of the system.
According to our calculations, this time is 
a few $\times$ 10,000 years.
Also, the shock wave material is injected preferentially in the outer
parts of the collapsing core, resulting in spatial gradients
in the distribution of the radioactivities.
This is evident in {\uppercase {Fig.}} 3, which shows the ratio between
the amount of shock wave material
with respect to the total accumulated mass as a function of distance
from the forming protostar.
The fractional shock mass shows a clear gradient toward the outer
parts of the nebula, and there are bumps in the gradient according
to the clumps of matter being injected by the RT-fingers at that
distance from the center of the system.
These results lead us to expect that the material
arrives into the forming solar system in the form of clumpy
infall instead of homogeneous rain of well-mixed material.
Since our calculations do not include
the rotation of the core, we cannot follow the formation of the protostellar
disk.  Therefore, it might be possible that the arriving material
would be mixed within the disk, but a preliminary study of mixing
in protostellar disks suggests that heterogeneities might survive
even at that stage
\citep{bos01}.


In addition to confirming the injection results of the previous calculations,
our results also support the original conclusion of \citet{fos96}
and \citet{van98} on the competition between the self-gravity of the core
and the development of instabilities at the contact surface between the
shock wave and the cloud.
In the case described here and in the previous simulations
of successful triggered collapse - cases of intermediate velocity
shock waves striking centrally condensed molecular cloud cores -
the self-gravity pulls the core to the point of collapse
before the instabilities have had a chance to develop fully.
At this point, the instabilities serve as feeders of shock wave material
into the forming solar system instead of destroying the core, as suggested
by calculations of high-velocity shock waves.

\section{Discussion}

Our calculations of the impact of an interstellar shock wave on
a molecular cloud core confirm the previous results:
in isothermal shock waves, shock wave material is injected into the
collapsing system by Rayleigh-Taylor instabilities developing at the
surface of the compressed core.  Injection begins shortly after the
central density of the core has started to grow due to the
shock wave transmitted from the compressed outer regions of the core.
The injection efficiency is $\sim$10\%,
and the amount of injected material in the central parts
of the collapsing core is typically $\sim$0.1\% of the total mass
contained in that region.

Our high-resolution calculations show that the RT-fingers retain
their structure down to the size of the forming solar system.  We therefore
expect the injected shock wave material to arrive on the disk as
clumpy infall rather than as thoroughly mixed material.
It is possible that instabilities might stir sufficient turbulence
in the infalling gas to cause the material to be thoroughly mixed before
it arrives on the disk, or that small-scale dynamically driven instabilities
could occur at the surface of the RT-fingers at a scale that is beyond the
resolution of our calculations.  However, we do not see any evidence for
this in our calculations, and our main conclusion therefore remains that
the arrival of the injected shock wave material into the protostellar disk
was neither spatially or temporally homogeneous.
Our calculations suggest that the total number of RT-fingers
reaching into the forming solar system is 10-12.  If the RT-fingers
remain separate on their way to the protostellar disk, the spacing between
the fingers as they strike the disk depends on the size of the disk.
For the 3000 AU of the panels of {\uppercase {Fig.}} 2
- the size of the forming disk -
the fingers would be roughly 300 AU apart, while a later-stage disk
of 40 AU, for example, would see a spacing of $\sim$4 AU of the RT-fingers
as they land on the disk.

\acknowledgments{
The calculations were performed on the Carnegie Alpha
Cluster, which is supported in part by NSF MRI grant AST-9976645.
This work was also supported by NASA Origins of Solar Systems Program
grant NAG5-4306 and NASA Astrophysical Theory Program grant NAG5-9263.
}

\clearpage

\begin{figure}
\plotone{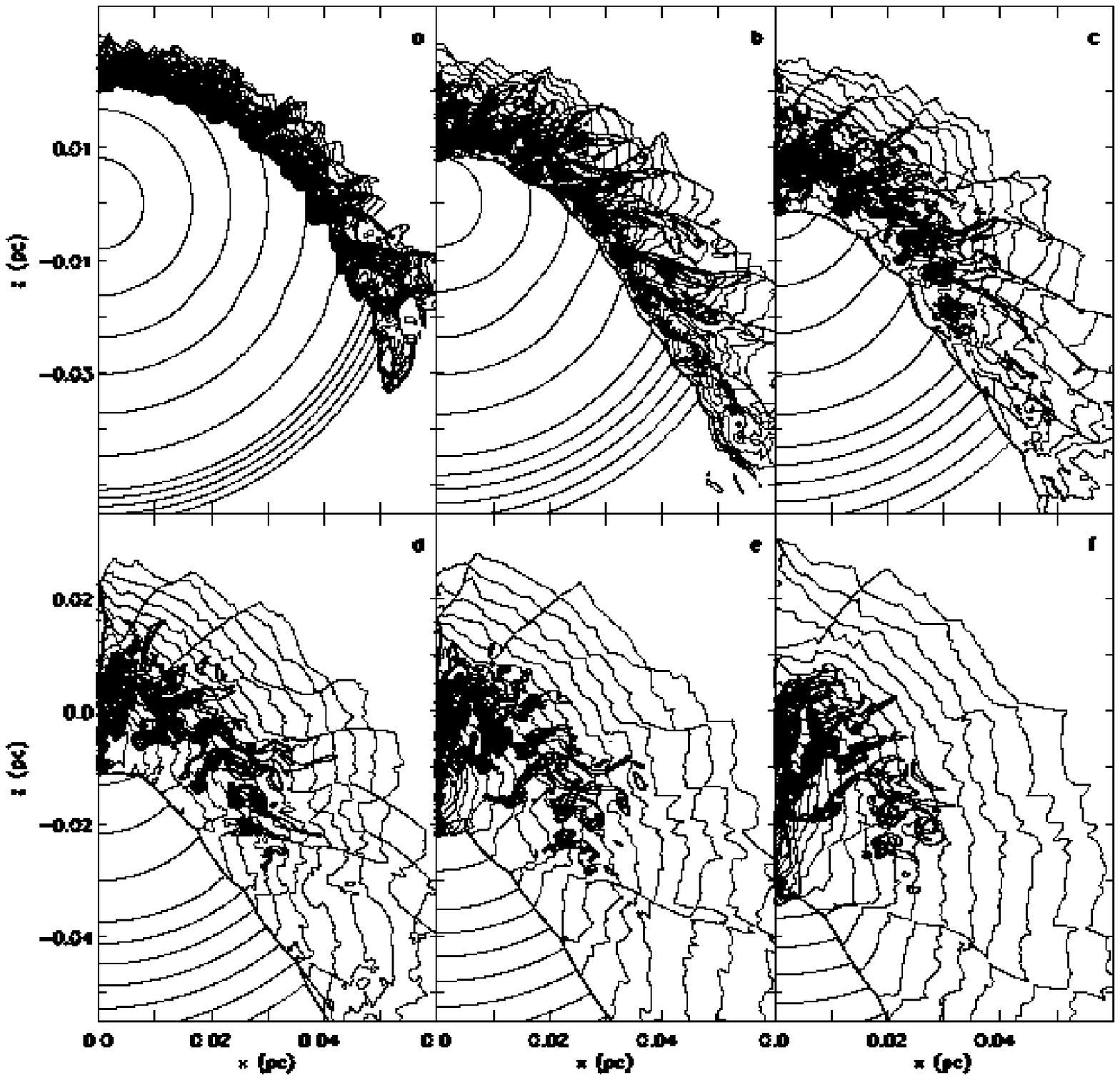}
\caption{
The development of Rayleigh-Taylor fingers.  The system is shown at $t$ =
22,000 yr (a), 44,000 yr (b), 66,000 yr (c), 88,000 yr (d),
110,000 yr (e), and 132,000 yr (f).
The thin contours depict the gas density in the system and range from
6.20$\times$10$^{-21}$ g cm$^{-3}$ (1/100 the initial central density) to
7.93$\times$10$^{-16}$ g cm$^{-3}$, with each contour representing
a change of factor 1.5 in density.
The thick contours show the behavior of the color field and range from
0 to 3.40$\times$10$^{-20}$ g cm$^{-3}$ in steps of
2.43$\times$10$^{-21}$ g cm$^{-3}$.
\label{fig01}}
\end{figure}

\begin{figure}
\plotone{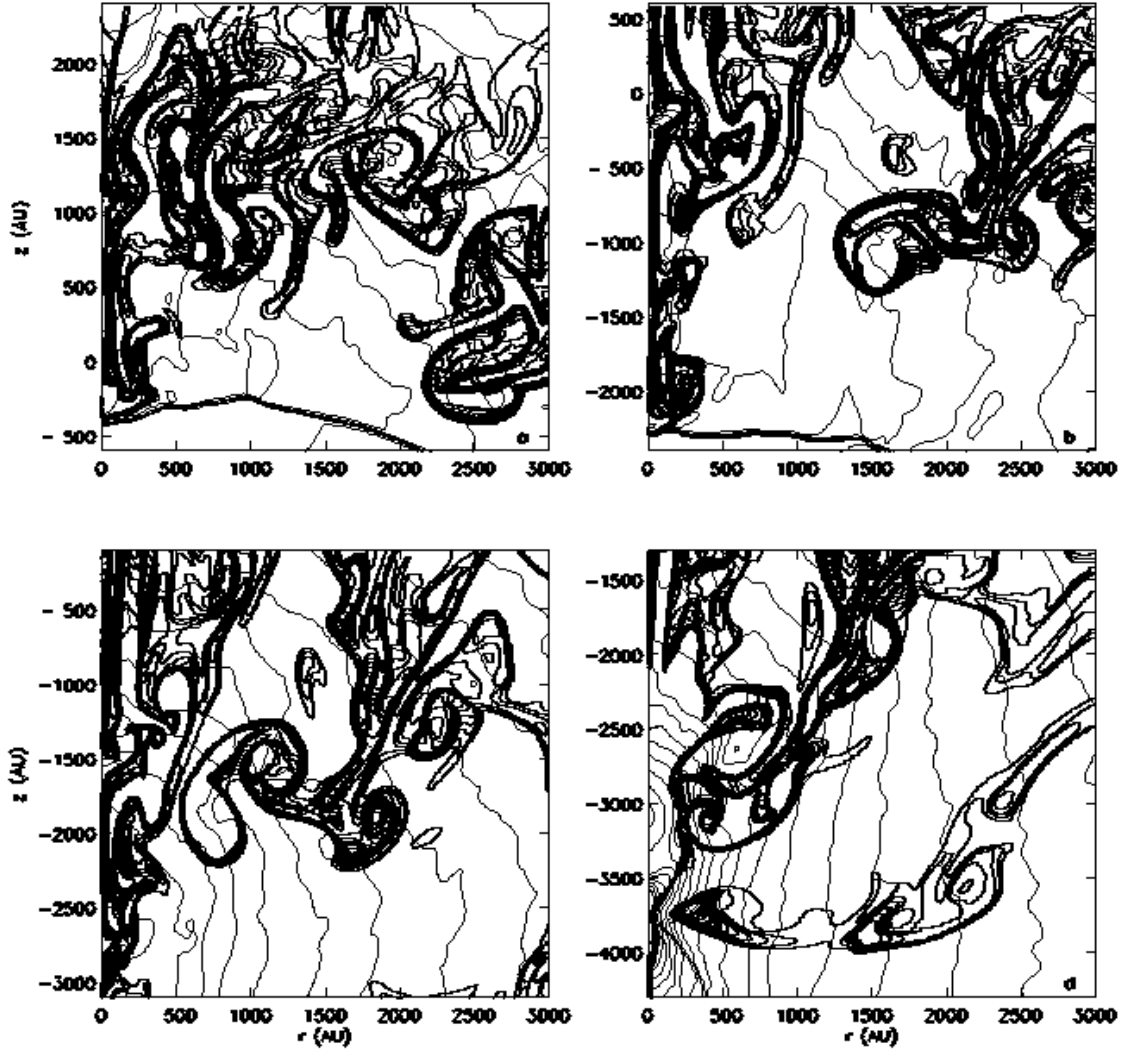}
\caption{
Closeup of the Rayleigh-Taylor fingers.  The system is shown at $t$ =
66,000 yr (a), 88,000 yr (b), 110,000 yr (c), and
132,000 yr (d).
The thin contours depict the gas density in the system and range from
6.20$\times$10$^{-21}$ g cm$^{-3}$ to
7.93$\times$10$^{-16}$ g cm$^{-3}$, with each contour representing
a change of factor 1.5 in density.
The thick contours show the behavior of the color field and range from
0 to 3.40$\times$10$^{-20}$ g cm$^{-3}$ in steps of
2.43$\times$10$^{-21}$ g cm$^{-3}$.
\label{fig02}}
\end{figure}

\begin{figure}
\plotone{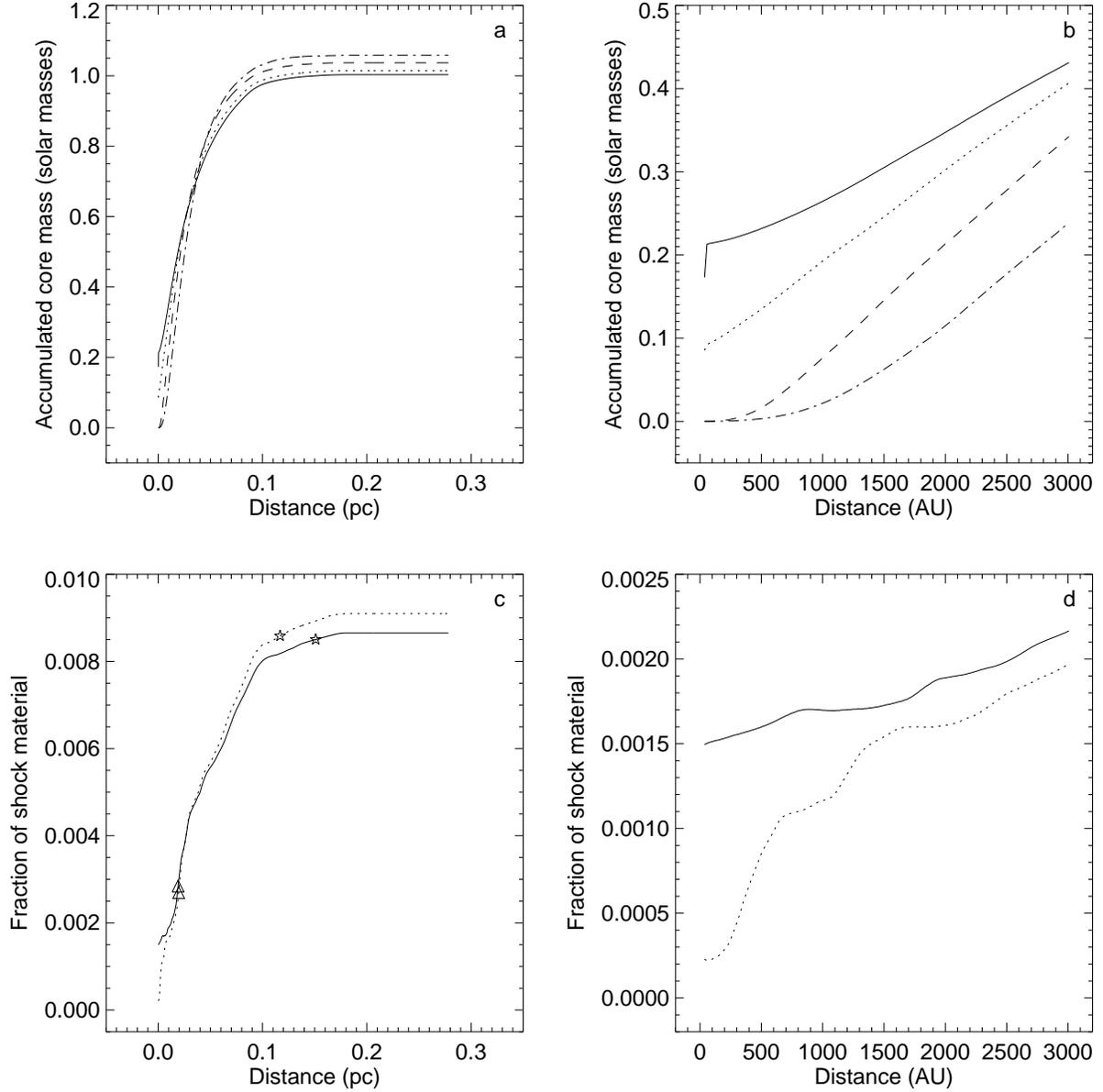}
\caption{
The total accumulated mass (a,b) and
the fraction of the shock wave material of the accumulated mass (c,d)
as a function of distance from the forming protostar.
The different lines correspond to the system at $t$ =
88,000 yr ({\it dashed-dotted line}),
110,000 yr ({\it dashed line}),
132,000 yr ({\it dotted line}), and
140,000 yr ({\it solid line}).
The right-hand panels (b,d) are a closeup of the inner 3000 AU
of the whole system shown in the left-hand panels (a,c).
In the bottom panels, the triangle and the star correspond to the distance
within which the accumulated core mass is 0.5 and 1.0 solar masses,
respectively.
\label{fig03}}
\end{figure}

\end{document}